\documentclass[a4paper,prl,twocolumn,showpacs,10pt,aps,reprint,superscriptaddress]{revtex4-1}
\usepackage{amsmath,amssymb,hyperref}
\usepackage{graphicx}
\usepackage{datetime}
\usepackage{braket}
\usepackage{xfrac}
\usepackage[usenames,dvipsnames]{color}
\usepackage[]{pdfpages}
\pdfoutput=1 
\makeatletter
\AtBeginDocument{\let\LS@rot\@undefined}
\makeatother

\definecolor{MyGreen}{rgb}{0,0.4,0}

\newcommand{\figref}[1]{Fig.~\ref{#1}}
\newcommand{\eqnref}[1]{Eqn.~(\ref{#1})}

\newcommand{\micron}{\ensuremath{\mu{\rm m}}}
\newcommand{\gOne}{\ensuremath{g^{(1)}}}
\newcommand{\ntot}{\ensuremath{n_{tot}}}

\newcommand{\supp}{Supplementary Material}

\newcommand{\Imperial}{
Quantum Optics and Laser Science group, Blackett Laboratory, Imperial College London, Prince Consort Road, SW7 2AZ, UK}
\newcommand{\ImperialCQD}{Centre for Doctoral Training in Controlled Quantum Dynamics, Imperial College London, Prince Consort Road, SW7 2AZ, UK}
\newcommand{\Oxford}{Department of Materials, University of Oxford, Parks Road, Oxford, OX1 3PH, UK}
\newcommand{\Karlsruhe}{Physikalisches Institut, Karlsruher Institut f\"ur Technologie, Wolfgang-Gaede-Stra\ss{}e 1, 76131 Karlsruhe, Germany}

\begin{document}

\title{Driven-dissipative, non-equilibrium Bose-Einstein condensation of just a few photons}

\author{Benjamin T. Walker}\affiliation{\Imperial}\affiliation{\ImperialCQD}
\author{Lucas C. Flatten}\affiliation{\Oxford}
\author{Henry J. Hesten}\affiliation{\Imperial}\affiliation{\ImperialCQD}
\author{Florian Mintert}\affiliation{\Imperial}
\author{David Hunger}\affiliation{\Karlsruhe}
\author{Aur\'elien A. P. Trichet}\affiliation{\Oxford}
\author{Jason M. Smith}\affiliation{\Oxford}
\author{Robert A. Nyman}\email[Correspondence to ]{r.nyman@imperial.ac.uk} \affiliation{\Imperial}

\date{\today}

\begin{abstract}
 \textbf{Coherence is a defining feature
of quantum condensates. These condensates are inherently multimode phenomena and in the macroscopic limit it becomes extremely difficult to resolve populations of individual modes and the coherence between them. In this work we demonstrate non-equilibrium Bose-Einstein condensation (BEC) of photons in a sculpted dye-filled microcavity, where threshold is found for $8\pm 2$ photons. With this nanocondensate we are able to measure occupancies and coherences of individual energy levels of the bosonic field. Coherence of individual modes generally increases with increasing photon number, but at the breakdown of thermal equilibrium we observe multimode-condensation phase transitions wherein coherence unexpectedly decreases with increasing population, suggesting that the photons show strong inter-mode phase or number correlations despite the absence of a direct nonlinearity. Experiments are well-matched to a detailed non-equilibrium model. We find that microlaser and Bose-Einstein statistics each describe complementary parts of our data and are limits of our model in appropriate regimes, which informs the debate on the differences between the two~\cite{Scully99, Miesner98}.
 }
\end{abstract}

\maketitle

%
%
While BEC is a general phenomenon, the standards of experimental evidence needed to claim BEC differ among different physical realisations. Ultracold atomic gases are very nearly closed systems for which a purely equilibrium description is often sufficient, so macroscopic occupancy of one state is considered proof of BEC. How condensation can be demonstrated with microscopic particle numbers is an open question. Quantum gases in pure states with as few as two fermions~\cite{Serwane11} or six bosons~\cite{Kaufman16} have been created in very specific configurations, but BEC is notoriously difficult to achieve by thermalisation in smooth traps~\cite{Chuu05, Bourgain13}.

For polariton condensates in microcavities it is now accepted that the build-up of coherence and population in lasing arise from stimulated emission~\cite{Deng02}, but in condensation the build-up is caused by bosonically-stimulated scattering among the polaritons. Despite the finite-number of particles, as few as about 100 in Ref.~\cite{Kasprzak06}, the question of how a threshold for BEC, a phase transition defined only in the thermodynamic limit, can be determined has rarely been considered.

The original thermodynamic-limit, equilibrium Penrose-Onsager criterion for BEC is that, as system size grows towards infinity, a finite fraction of particles remain found in the lowest energy mode~\cite{Penrose56}. More general definitions of condensation in multimode systems have recently come into question, applicable not only to bosonic statistics out of equilibrium~\cite{Schnell17} but also to networks, traffic jams, evolutionary game theory, population dynamics and chemical reaction kinetics~\cite{Knebel15}. Condensation is said to have occurred when the populations of some modes of the system grow linearly with total population as the total tends to infinity, while other modes are depleted, with sub-linear growth or saturation. BEC is the special case where the only mode to condense is the lowest-energy mode. These clear theoretical definitions are not applicable to experiments, which cannot reach infinite populations.

An operational criterion for condensation applicable to experiments would consist of an inequality which defines a parameter region of condensation. Distinctions between true thermal equilibrium and near- or non-equilibrium situations call for robust criteria, which we will define in this manuscript. BEC can be distinguished from general or multimode condensation in that only one mode, the lowest-energy mode, is condensed, and all other modes depleted.

In this work, we optically pump a fluorescent dye in an open microcavity consisting of a planar mirror and a microfabricated, concave mirror, an exemplary open BEC system~\cite{Klaers10b, Marelic15}. Through incoherent emission and re-absorption and dye ro-vibrational relaxation, excitations are exchanged between dye molecules and cavity photons, and the photons can reach thermal equilibrium near room temperature~\cite{Kirton15}. We use microfabricated mirrors to achieve large mode spacings, with spectroscopic resolution of the individual energy levels for the bosonic field: see \figref{fig: setup and BEC}(a).

In dye-microcavity photon BEC, thermalisation among the particles is, uniquely for quantum fluids, completely negligible as interactions are weak~\cite{Nyman14, vanderWurff14, Dung17}. Condensation is distinguished from lasing by thermalisation through multiple re-absorption and emission events for photons before they leave the microcavity~\cite{Schmitt15}. Either a good fit to the Bose-Einstein distribution~\cite{Klaers10b}, or the robustness of the lowest-energy mode as the strongly populated mode~\cite{Hesten18} are considered proof of BEC.


\subsection*{Demonstration of BEC of just a few photons}

\begin{figure*}[ht!]
  \includegraphics[width=0.32\textwidth]{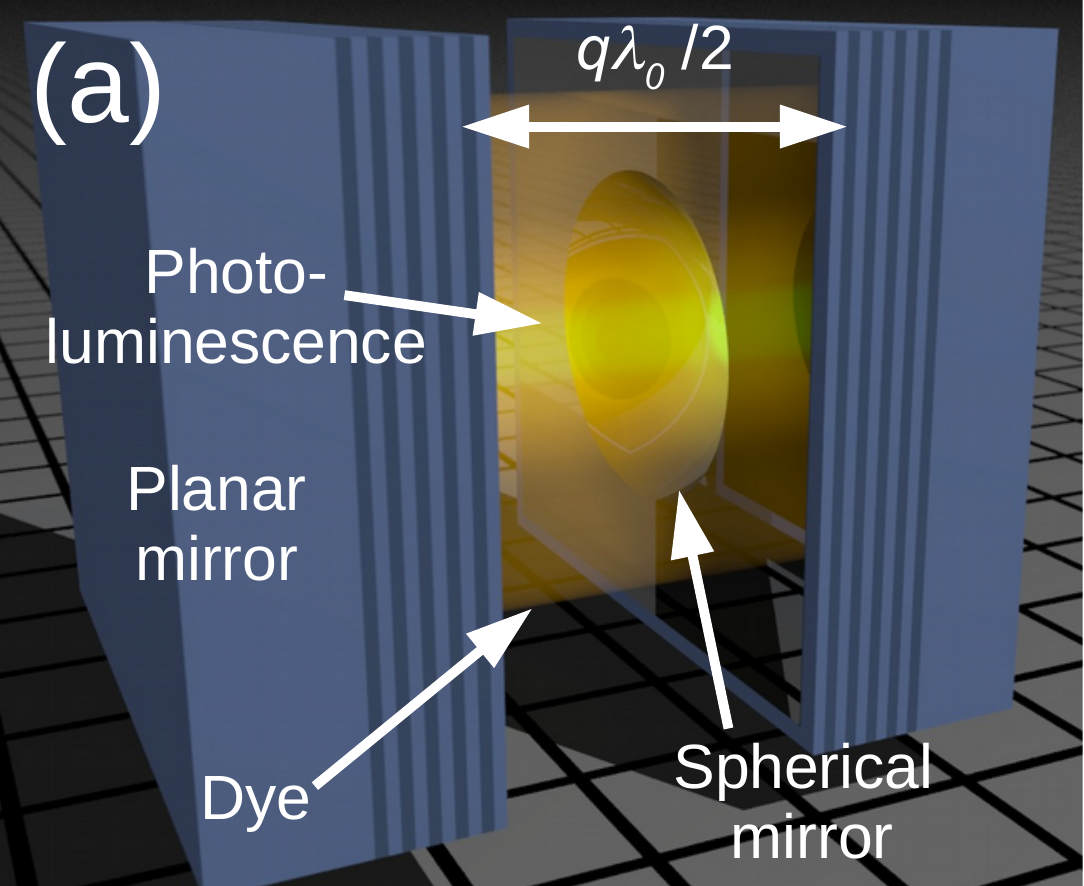}
  \includegraphics[width=0.38\textwidth]{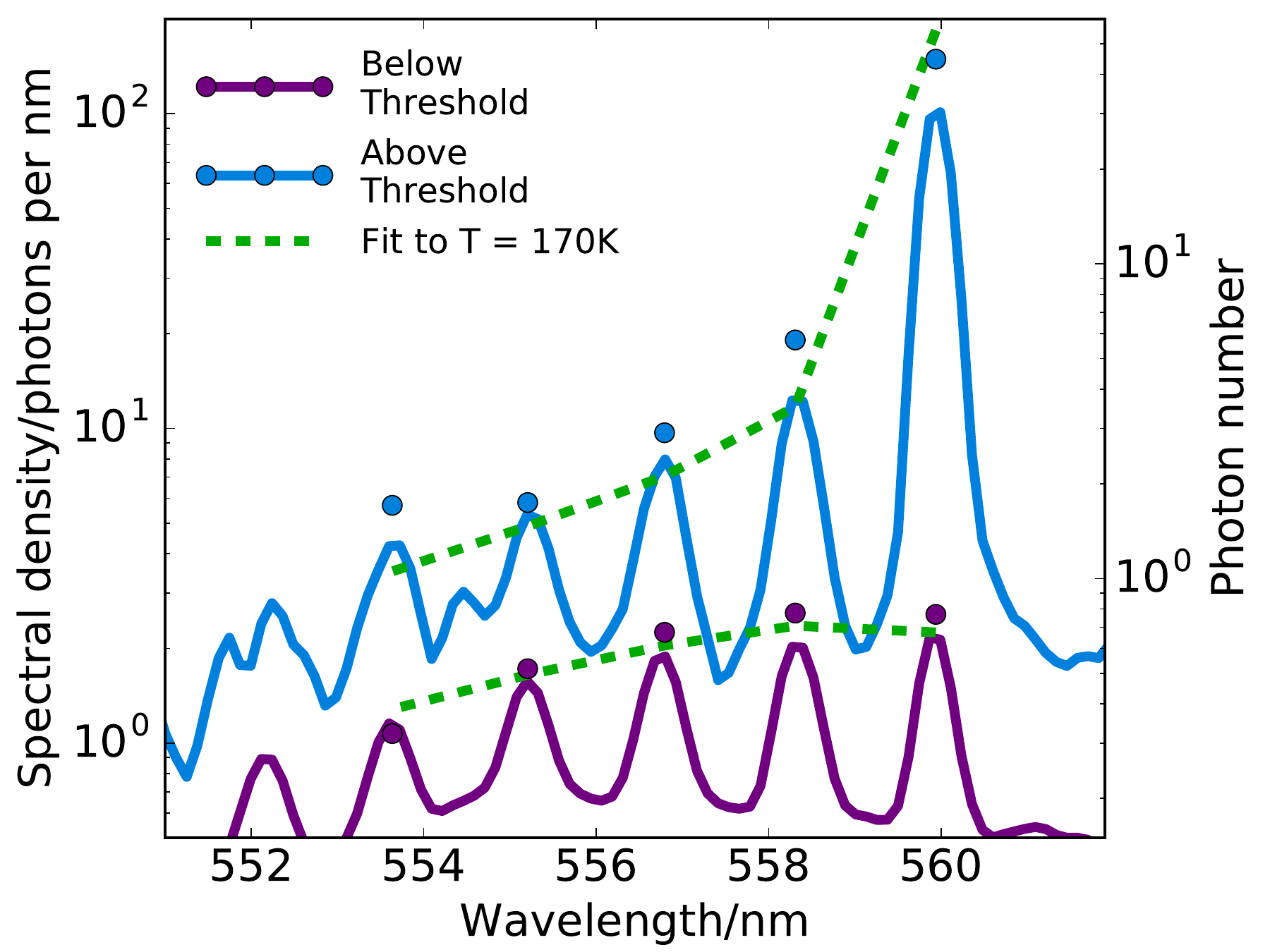}
  \includegraphics[width=0.70\textwidth]{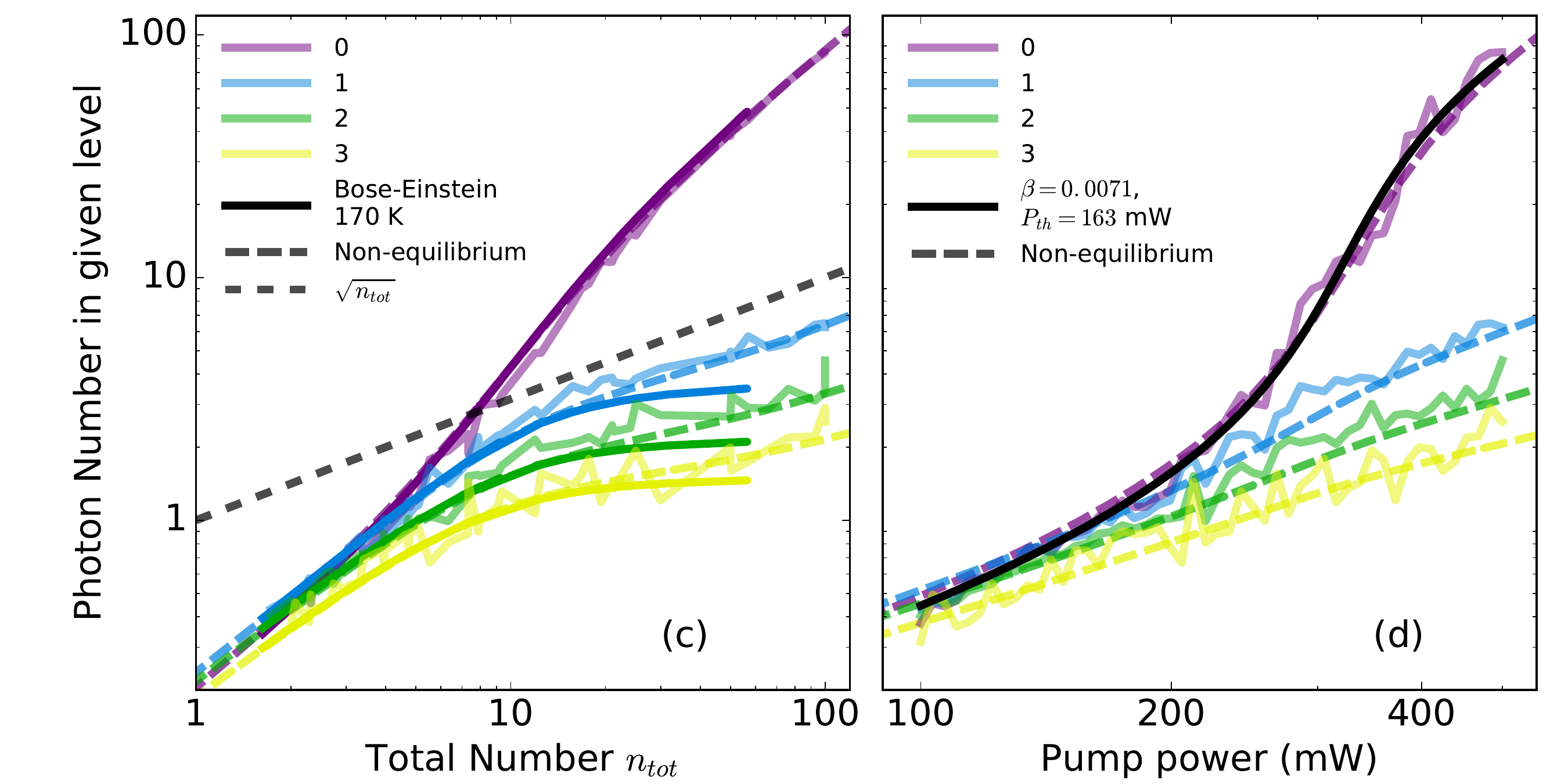}
  \caption{
  (a) The microcavity used for photon thermalisation and condensation, of length $q \lambda /2$ where $\lambda$ is the wavelength of light in the medium, and $q$ the longitudinal mode number. The experiments here have $q$ in the range 9 to 11. (b) Sample photoluminescence spectra showing that individual energy levels are easily resolved and can be assigned photon populations (shown as dots). The decay of population for higher energy levels is indicative of a thermal distribution. Threshold behaviour is shown in (c) and (d), with energy level labelled in the legend. As a function of total photon number (c), the population distribution can fitted with the Bose-Einstein distribution of \eqnref{eqn: BE distn}, revealing a broadened threshold at $8\pm 2$~photons on average, accompanied by near saturation of excited state populations. A simple microlaser model (solid black line in panel (d)) is more appropriate when considering the lowest-energy level mode as a function of nominal pump power, revealing the fraction $\beta$ of spontaneous emission into the cavity ground-state mode and a threshold at $P_{th}$. Dashed lines in panels (c) and (d) are results of a multi-mode non-equilibrium simulation, whose main adjustable parameter is the pump spot size, set to 1.2~\micron. 
  }
  \label{fig: setup and BEC}
\end{figure*}

Photon BEC has previously been reported with as few as 70 photons~\cite{Dung17}. In this work, we have achieved photon thermalisation and BEC with just 8 photons, arguably the smallest BEC ever published. Our concave mirror radius of curvature of 400~\micron, defines a two-dimensional harmonic oscillator (2DHO) potential of frequency $f=1.4$--$1.7$~THz (level spacing $\epsilon = h f$) depending on the longitudinal mode number $q$ (we use $9\leq q \leq 11$). By varying the cavity length, we set the energy of the lowest level, equivalent to a cutoff wavelength $\lambda_0$, between 555 and 580~nm: see \figref{fig: setup and BEC}(b). We observe cavity photoluminescence with a spectrometer of resolution 0.3~nm (equivalent to $0.3$~THz), sufficient to distinguish individual energy levels.

\figref{fig: setup and BEC}(c) shows the population of each of the lowest 4 energy levels as a function of total population (bottom left) or pump power (bottom right), for a single data set. Threshold is a broad feature, characteristic of finite-sized systems. We compare the populations of all modes to a simple thermal equilibrium Bose-Einstein distribution and to a full non-equilibrium model~\cite{Kirton15, Keeling16}, as well as a single-mode microlaser model for the lowest energy level. In the \supp\ we show how the equilibrium and microlaser models can be derived as limiting cases of the full model. The thermal equilibrium model (panel (c), solid lines) uses Bose-Einstein statistics so the population $n_i$ of the $i$th excited mode is 
\begin{align}
 n_i = \frac{g_i}{{\rm e}^{(\epsilon_i-\mu)/k_B T}-1}\label{eqn: BE distn}
\end{align}
with the degeneracy $g_i=i+1$ for a single spin state of a 2DHO, the mode energy is $\epsilon_i = i\,hf$, the typical thermal energy is $k_B T$, and $\mu$ the chemical potential determines \ntot. The least-squares fit returns $T = 170\pm 20$~K, somewhat below room temperature which is a consequence of the fact that the system is not quite at equilibrium. 

The non-equilibrium model has three adjustable parameters: the pump-spot size, the rate of spontaneous emission into free space and the calibration of the detection system in terms of photon number (see \supp\ for details). Notably the spontaneous emission rate is reduced significantly from its free-space value, since most emitted light is re-captured by the cavity mirrors.  
The populations of excited levels nearly saturate above threshold (as they would for exact equilibrium), a feature which is well described by the non-equilibrium model.

We have measured through a linear polariser aligned to maximise transmission of the condensate light, to avoid ambiguities in the role of polarisation, which recent results indicate will not affect our conclusions~\cite{Moodie17, Greveling17} (see also \supp\ for details).

Four suitable criteria for condensation, based on mode populations $n_i$ and the total population $\ntot = \sum_i n_i$, where the indices $i$ run over all modes which can be measured or calculated are (i) $n_0 > \ntot/2$~\cite{Schnell17}; (ii) $\ntot > \lim_{\ntot \rightarrow \infty} \left\{\ntot - n_0 \right\}$~\cite{Mullin97}; (iii) $n_i > k_BT/\epsilon$ with $T$ being the effective temperature~\cite{Kirton15}; and (iv) $n_i > \ntot^{1/\alpha}$ where $\alpha$ is the dimensionality of the system (at least unity). These concepts are discussed in more detail in the \supp.
Criteria (ii) and (iii) are derived in near-equilibrium conditions, and (i) is very strict, forcing single-mode condensation. 
%
%
Thus, (iv) is the only criterion which is also applicable to multimode condensation which is known to occur in photon condensates~\cite{Hesten18, Marelic16a}. It would also be useful in categorising other non-equilibrium condensation processes, such as prethermalization~\cite{Gring12}.

The dimensionality of this system is $\alpha = 2$, so criterion (iv) for condensation in mode $i$ becomes $n_i > \sqrt{\ntot}$. The mirror shape defines an effective 2DHO potential, for which the critical total particle number (ii) becomes $\ntot > (\pi^2/6)(k_B T / \epsilon)^2$ for level spacing $\epsilon$. In \figref{fig: setup and BEC}(c) the dashed line illustrates criterion (iv). Criteria (ii)--(iv) nearly coincide and yield a threshold of $\ntot = 8\pm 2$ photons. Criterion (i) contradicts these, requiring not only that condensation be found but also that multimode condensation be excluded, and gives $\ntot = 16$ photons for BEC. Even at $8\pm 2$ photons, BEC is well established.


\subsection*{Breakdown of thermal equilibrium by multimode condensation}


For the conditions of \figref{fig: setup and BEC} absorption events happen 4 times faster than cavity loss, so photons can exchange energy with the thermal bath of dye-solvent vibrations, and thermal equilibrium and BEC are good descriptions.

\begin{figure*}[hbt!]
  \includegraphics[width=0.8\textwidth]{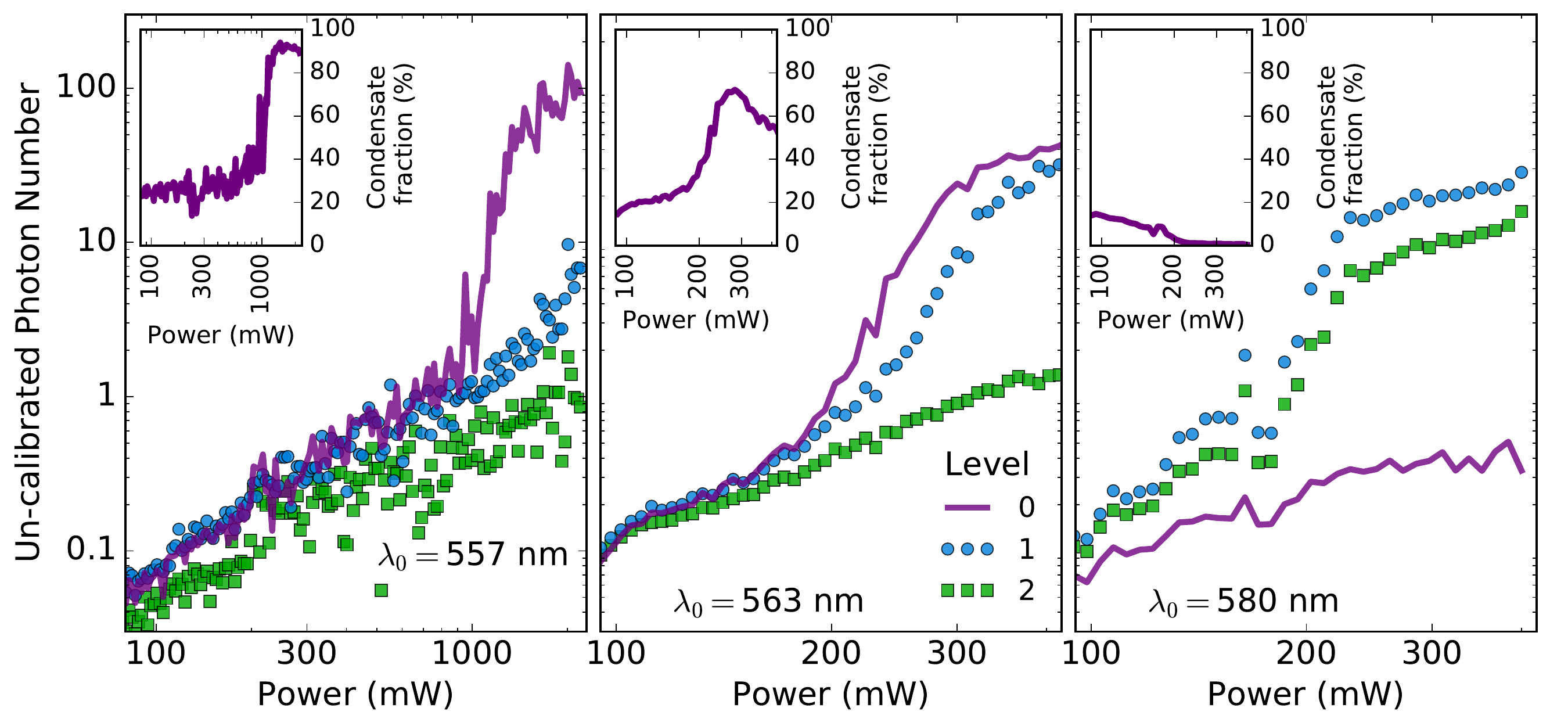}
  \includegraphics[width=0.83\textwidth]{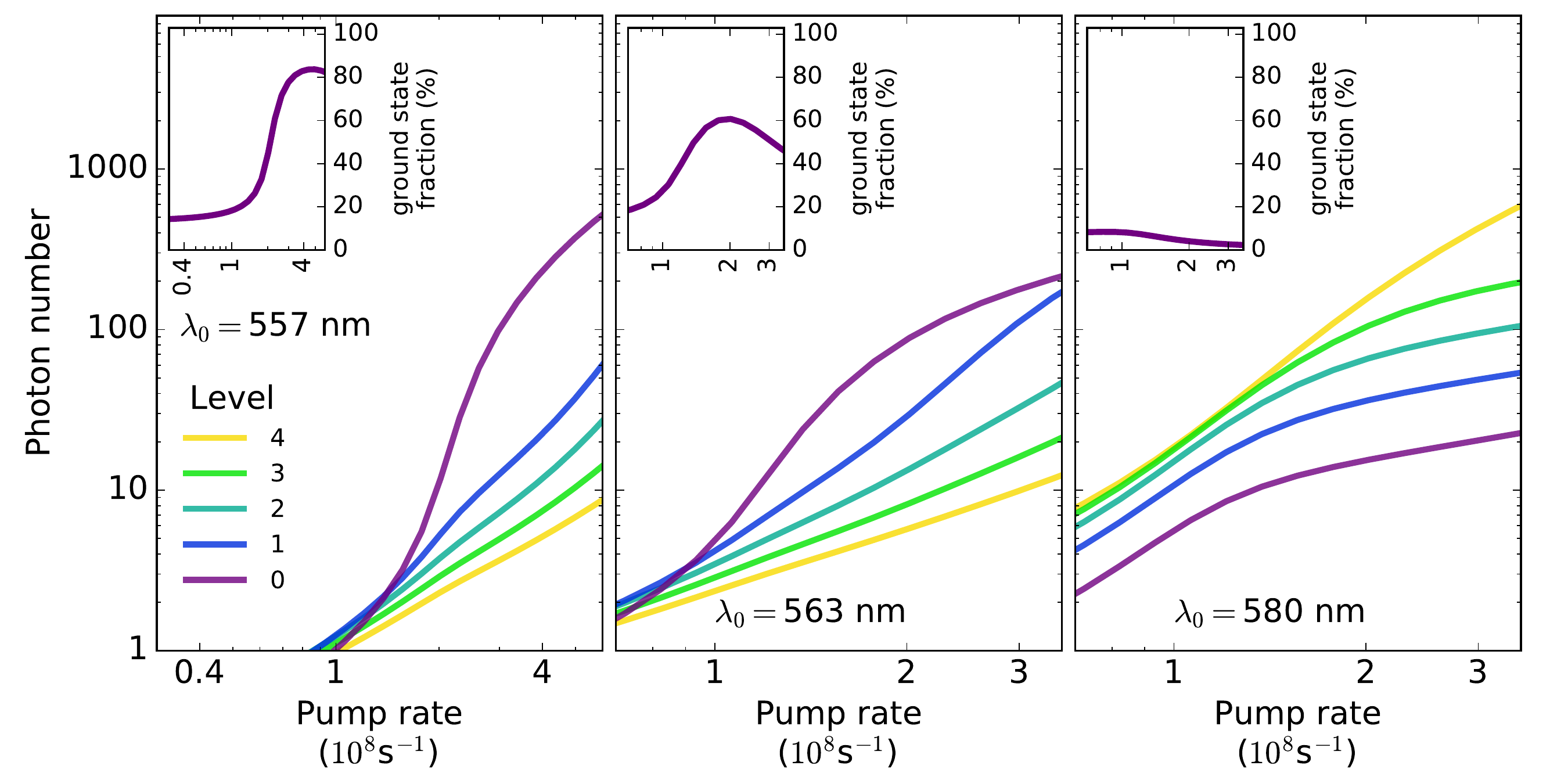}
  \caption{The breakdown of thermalisation, comparing experiments (upper row, number uncalibrated) to a non-equilibrium model (lower row). Rapid thermalisation through re-absorption (left panels, with cutoff wavelength $\lambda_0 = 557$~nm) leads to Bose-Einstein condensation meaning large population of the ground state accompanied by saturation of excited-state populations. When thermalisation through photon re-absorption is no faster than cavity loss, multiple modes condense (middle panel, $\lambda_0 = 563$~nm). For extremely weak re-absorption, lasing can occur in any mode or modes, not necessarily including the ground state (right panel, $\lambda_0 = 580$~nm). The only adjustable parameter in the model is the pump spot size, set to 2.4~\micron. The average numbers of re-absorption events per cavity-loss time are 6.7, 2.7 and 0.15 for $\lambda_0 = $557, 563 and 580~nm respectively.}
  \label{fig: thermalisation and breakdown multimode}
\end{figure*}


In \figref{fig: thermalisation and breakdown multimode} the effect of reducing the rate of thermalisation through re-absorption is shown. The re-absorption rate is $\overline{n}_{mol} \sigma(\lambda)c^*$ with $\overline{n}_{mol}$ the effective molecular number density (see \supp\ for details), $c^*$ the speed of light in the medium and $\sigma(\lambda)$ the absorption cross-section at wavelength $\lambda$. The degree of thermalisation is parameterised by the ratio of re-absorption to cavity loss rates: $\gamma = \overline{n}_{mol} \sigma(\lambda)c^* / \kappa$. We estimate (see below) that $\kappa = 5$~ps.
Experiments (upper row) are compared with the full non-equilibrium model (lower row), with the same parameters as \figref{fig: setup and BEC} except the pump spot size, set to 2.4~\micron. In the left panels, the degree of thermalisation is $\gamma = 6.7$, so the system is strongly thermalised and a condensation threshold is reached for the lowest energy mode, and no other level. Up to 95\% of photons are in this nearly pure BEC. 

For weaker thermalisation (centre panels, $\gamma = 2.7$), the lowest-energy level shows threshold, but one or more excited levels also show threshold, and the lowest-energy--level fraction peaks around 75\%. Multi-mode condensation occurs at higher pump powers.
For very weak re-absorption (right panels, $\gamma = 0.15$), multiple modes not including the ground state show threshold, and the system cannot even approximately be described as a BEC.

\subsection*{Coherence}

Having established that the near-thermalised photon population can be described by either Bose-Einstein statistics or a non-equilibrium model, we now apply these descriptions to the phase coherence, \gOne. We measure \gOne\ using a spectrometer on the output of a Mach-Zehnder interferometer, where we can delay one interferometer arm by some time $\tau$, as in Ref.~\cite{Marelic16a}.

\begin{figure}[ht!]
  \includegraphics[width=0.48\textwidth]{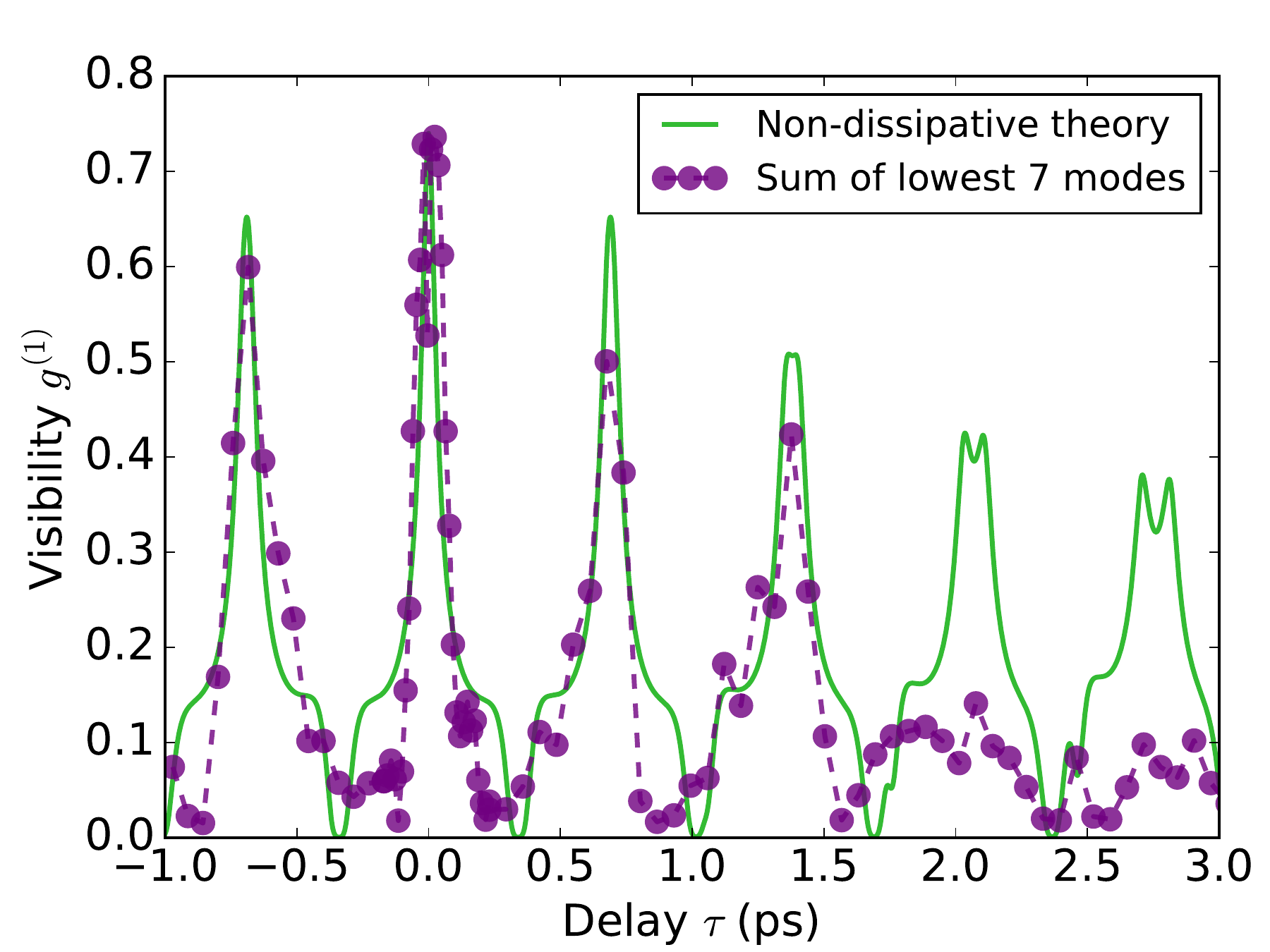}
  \includegraphics[width=0.48\textwidth]{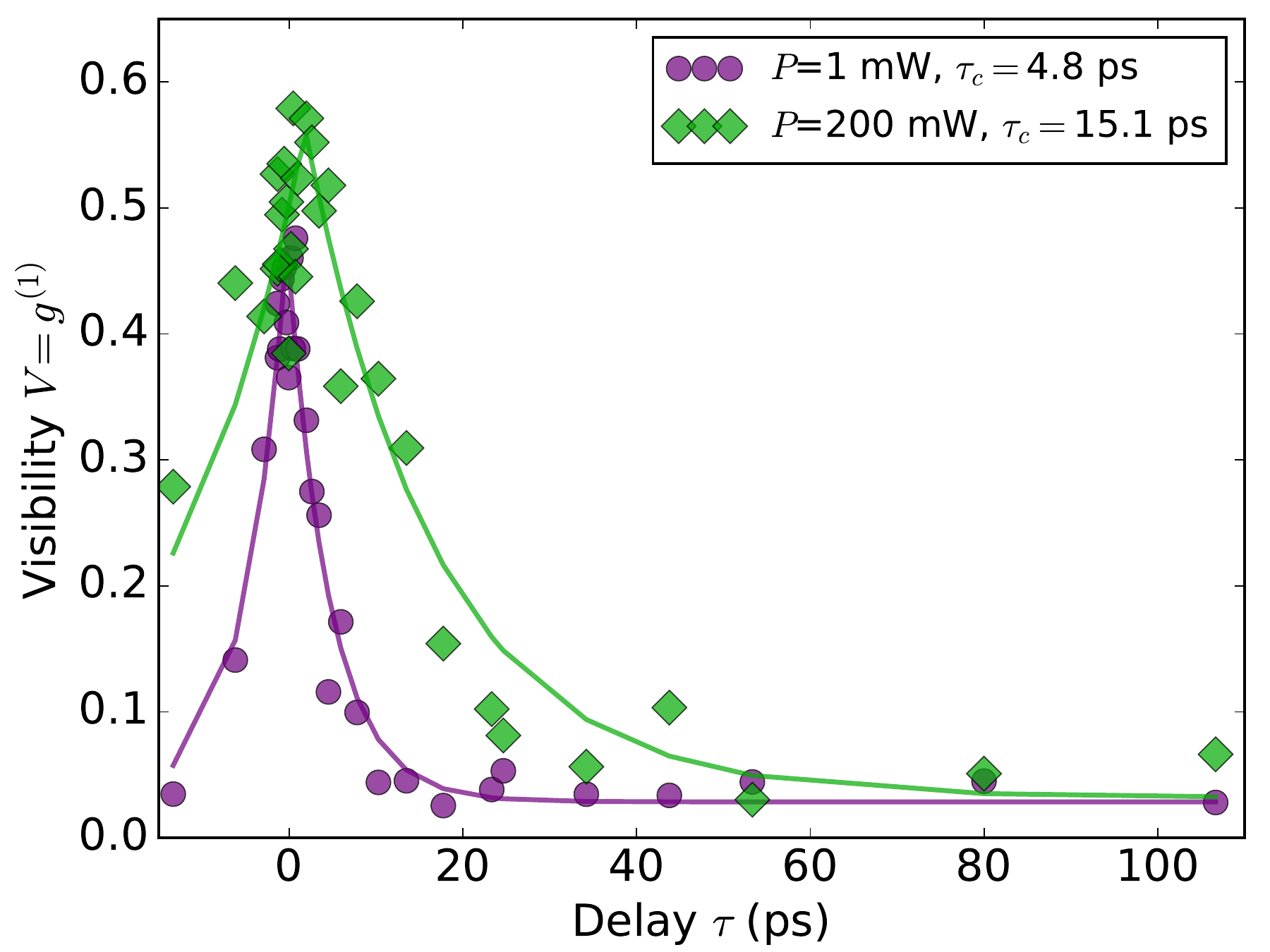}
  \caption{
    Phase coherence \gOne\ measured through a Mach-Zehnder interferometer using a spectrometer, for various delay times $\tau$. (top) Visibility is determined after summing signals from many modes. \gOne\ decays on thermal timescales similar to $h / k_B T$, then revives once every trap oscillation period. The trap frequencies for the two axes are 1.42 and 1.48~THz. Here, coherence is well described by closed-system Bose-Einstein statistics (solid line, as discussed in the main text). (bottom) Visibility determined for a single mode, the ground state. An exponential $\gOne(\tau) \propto \exp{(-|\tau-\tau_0|/\tau_c)}$ of coherence time $\tau_c$ fits the data well, which is typical of driven-dissipative systems like microlasers.
  }
  \label{fig: coherence both kinds}
\end{figure}

For a non-dissipative thermal Bose gas below condensation threshold number, $\gOne(\tau)$ decays as a Gaussian with a characteristic time of order $h / k_B T \simeq 100~{\rm fs}$\cite{Damm17}. It is predicted that revivals of all correlation functions will occur at intervals of the oscillation period~\cite{Kohnen15} as a consequence of uniform energy-level spacing. They are diminished by a slight anisotropy of the mirrors. Both effects, decay and revival, can be seen in \figref{fig: coherence both kinds} (top panel). \gOne\ is inferred from the fringe visibility after summing the signals of several modes, which cover almost all of the population.  The theory plotted is based on Ref.~\cite{Kohnen15} which makes use of a decomposition of the photon field-annihilation operator in a basis of the trap states. Taking a density operator which describes an equilibrium distribution at room temperature with energy spacings $h\times 1.42$ and $1.48$~THz for the two axes, we then calculate the expectation of the field-field correlations. Fluctuations of the photon field are propagating back and forth across the trapping potential as weakly-damped wavepackets.

The decay of revivals is in part due to dissipation. In \figref{fig: coherence both kinds} (bottom panel) we show the coherence for the lowest-energy level alone, without summing with other modes before inferring visibility. $\gOne(\tau)$ fits well to an exponential decay with a coherence time $\tau_c$, for a variety of parameters. 

\begin{figure}[t!]
  \includegraphics[width=0.46\textwidth]{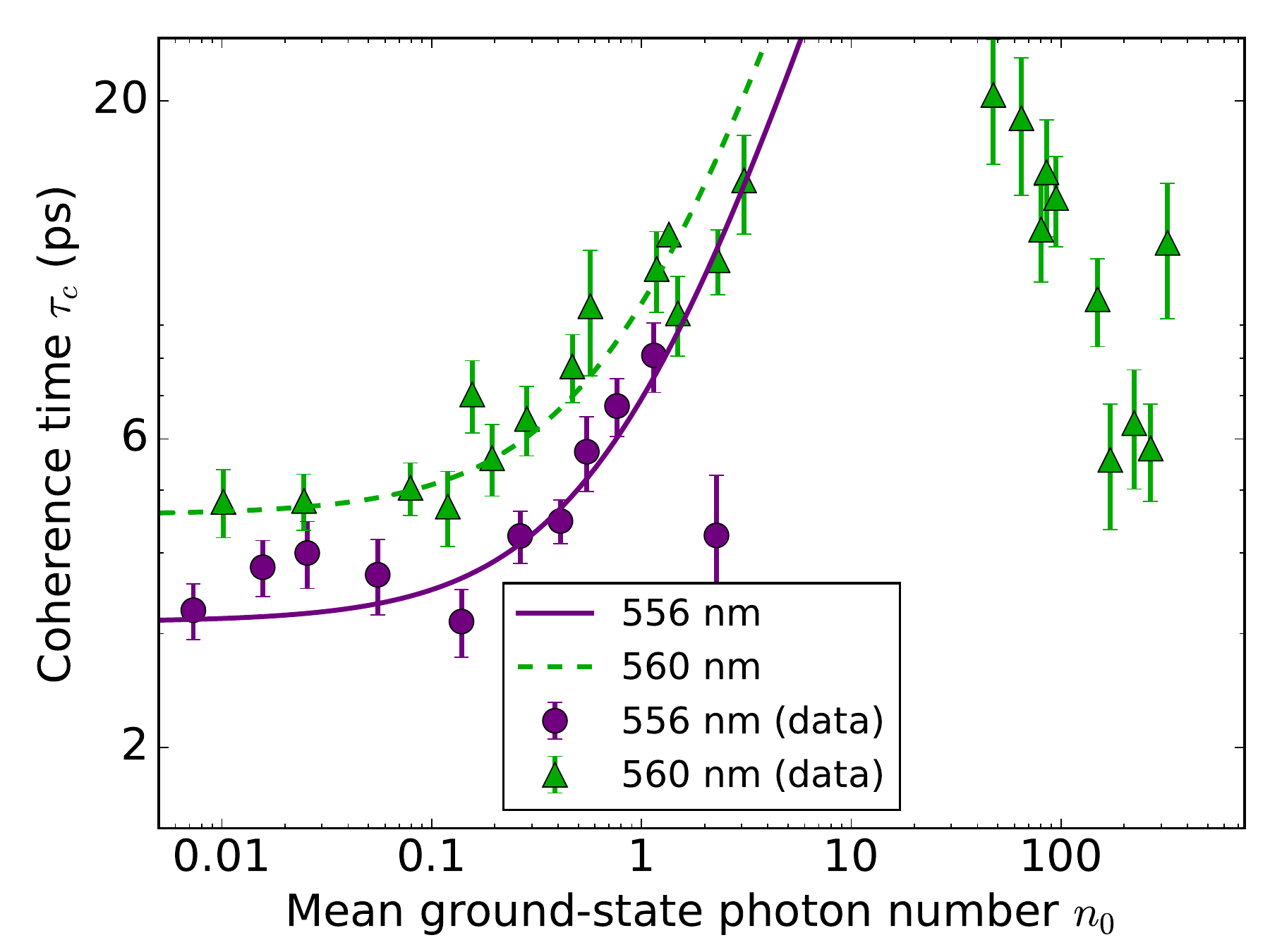}
  \includegraphics[width=0.46\textwidth]{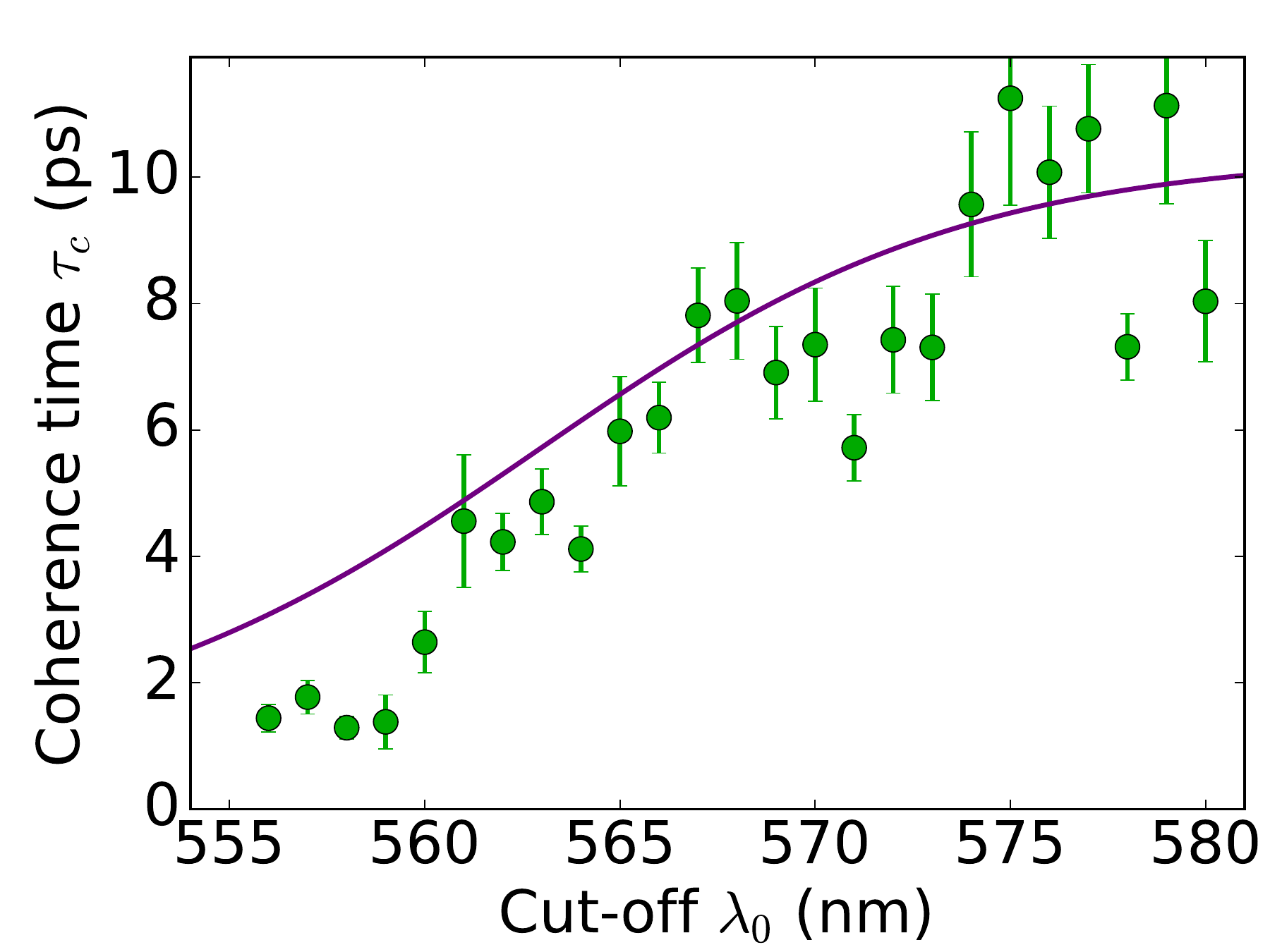}
  \caption{
  Coherence time $\tau_c$ of the ground state. Top: $\tau$ as a function of $n_0$ the population of the mode. Bottom: as a function of cavity cutoff wavelength $\lambda_0$ for small photon numbers $n_0<0.05$. Coherence time is independent of photon number for $n\ll 1$, but depends on the dissipation timescale, governed by both cavity loss $\kappa$ and re-absorption, the latter of which varies with $\lambda_0$. The only free parameter in the model is $1/\kappa = 5.2\pm 0.8$~ps. For increasing $n_0$ the coherence time increases, but for very large $n_0$ there is a dramatic and unexpected decrease in $\tau_c$. Uncertainties from least-squares fitting temporal coherence with exponential decays are represented by error bars. 
  }
  \label{fig: coherence single mode}
\end{figure}

By treating just one cavity mode, and assuming photon and molecule states are separable, one can reach a closed form for coherence time~\cite{Kirton15}: see \supp. For large photon numbers $n\gg 1$ (above threshold), the coherence time $\tau_c \propto n$, yielding the Schawlow-Townes limit. For $n\ll 1$, $\tau_c$ is independent of $n$, given by
\begin{align}
 \frac{1}{\tau_c} = \frac{1}{2}\left[ \kappa +  \overline{n}_{mol} \sigma(\lambda) c^*\right].
\end{align}
Coherence decays with half the rate at which photons are removed from the mode, both by cavity loss and by re-absorption, which in turn is the thermalisation rate. In \figref{fig: coherence single mode} we see quantitative agreement between experiment and theory for most parameters. The theory has only one adjustable parameter, the cavity-loss rate, for which we find $1/\kappa = 5.2\pm 0.8$~ps, in agreement with the value obtained through observations of the breakdown of the BEC description, \figref{fig: thermalisation and breakdown multimode}.

%
For very large photon numbers, $n\gtrsim 50$, coherence time decreases markedly with increasing photon number, in direct contradiction to the single-mode theory. This might be attributed to a breakdown of photon-molecule separability, or to inhomogeneous coupling of multiple modes to the molecular excitations~\cite{Hesten18}. 

\subsection*{Discussion}


The first-order coherence of the light can be interpreted through two complementary physical models: as a conservative thermalised Bose gas when accounting for many energy levels, or as a driven-dissipative open quantum system when inspecting the coherence of the lowest-energy mode alone. As an open quantum system, the coherence is limited by the re-absorption of the light, which is the very mechanism which induces the coherence-enhancing Bose-Einstein condensation itself. The tension between coherence and decoherence resolves at large photon numbers by a dramatically reduced coherence time, accompanied by multimode condensation.

From this extra decoherence mechanism, we infer that there is an effective interaction which couples the quantum states of the light across the multiple cavity modes, mediated by the dye molecules. This is despite the small measured value of the direct optical nonlinearity~\cite{Klaers10b}. Additionally, in a multimode condensate, photons in one condensed mode could act as reservoirs of excitations for other modes, enhancing number fluctuations and hence decreasing phase correlations~\cite{Marconi18}. Through this mechanism we anticipate that higher-order coherences such as inter-mode number correlations will lead to non-classical states of light, possibly including number squeezing. If such states can be understood and observed, they may well prove a valuable resource for quantum metrology as well as a fascinating subject of study in their own right, enabled by our microfabricated mirrors and photon thermalisation techniques.



\section*{References}
\bibliographystyle{prsty}
\bibliography{photon_bec_refs}

\section*{Acknowledgements}

We thank Rupert Oulton for enlightening discussions. We are grateful to the UK Engineering and Physical Sciences Research Council for supporting this work through fellowship EP/J017027/1 (R.A.N.) and the Controlled Quantum Dynamics CDT EP/L016524/1 (B.T.W. and H.J.H.). D.H. thanks the DFG cluster of excellence ``Nanosystems Initiative Munich''. L.C.F., A.A.P.T and J.M.S. acknowledge support from the Leverhulme Trust. 

\section*{Contributions:} B.T.W. carried out the experiments with assistance from R.A.N, and both analysed the data. L.C.F., A.A.P.T, J.M.S. and D.H. fabricated the mirrors and assessed their performance. H.J.H. and R.A.N. and worked out the theory with assistance from F.M. R.A.N. conceived the experiment, and wrote the manuscript with input from all authors.

\section*{Competing interests:} The authors declare no competing financial
interests.



\section*{Supplementary material}

# Driven-dissipative, non-equilibrium Bose-Einstein condensation of just a few photons: Supplementary material

Benjamin T. Walker,[1,2] Lucas C. Flatten,[3] Henry J. Hesten,[1,2] Florian Mintert,[1]
David Hunger,[4] Aurélien A. P. Trichet,[3] Jason M. Smith,[3] and Robert A. Nyman[1,*]

[1]*Quantum Optics and Laser Science group, Blackett Laboratory, Imperial College London, Prince Consort Road, SW7 2AZ, UK*
[2]*Centre for Doctoral Training in Controlled Quantum Dynamics, Imperial College London, Prince Consort Road, SW7 2AZ, UK*
[3]*Department of Materials, University of Oxford, Parks Road, Oxford, OX1 3PH, UK*
[4]*Physikalisches Institut, Karlsruher Institut für Technologie, Wolfgang-Gaede-Straße 1, 76131 Karlsruhe, Germany*

This document is intended as supplementary material, to be read alongside the Main Text.

## EXPERIMENTAL SYSTEM

Our experimental setup is largely as described in our previous articles [1–3], with an open microcavity using one planar mirror and one curved mirror, and the space between is filled with fluorescent dye (Rhodamine 6G in ethylene glycol). For these experiments, the radius of curvature (RoC) for the curved mirror was 400 $\mu m$, which we achieved by machining a superpolished substrate with focussed-ion-beam (FIB) milling [4–6] followed by smoothing by laser-induced heating [7] and commercial ion-beam sputtered coating of dielectric mirrors.

To prevent triplet-state population, we use 350 ns pump pulses, and as we vary pump laser power we retain near-constant time-averaged power by also changing the repetition rate. We record the calibrated spectrum of light emitted from the planar mirror and infer the populations of individual energy levels.

Light leaks from both sides of the microcavity. We image from the planar-mirror side onto a monitoring spectrometer. The response of this spectrometer to intracavity light is calibrated by comparison to the response to a known light power, together with the measured transmission of the mirrors. From the microfabricated mirror side, light is sent to a beamsplitter, with some directed to a camera, and most to an interferometer. The controls and data analysis techniques used with the interferometer are detailed in Ref. [2]. For this work, one output of the interferometer goes to a camera and the other quadrature goes to a spectrometer. We use this spectrometer for inferring coherence $g^{(1)}$.

## MIRROR FABRICATION

FIB milling makes nanometric precision of surface topography possible over a large range of feature sizes and radii of curvature. After the FIB milling, the surface of the concave feature was treated with a $CO_2$ laser to melt a thin layer of material and achieve a lower surface roughness upon resolidification [8]. Surface roughness was measured to be 0.3 nm (rms) after the smoothing step. See Fig. 1(a), where the right two columns were treated with the $CO_2$ laser and the left columns untreated.

To verify the optical quality of the concave surface the decay of the photon population of a longitudinal cavity mode after a short ($\approx 6$ ps) populating pulse was probed and a finesse of $2$–$5 \times 10^4$ were found, varying from feature to feature. At the 10th longitudinal cavity mode this finesse corresponds to a cavity lifetime of up to 150 ps. The nominal transmission and absorption for each dielectric mirror are 30 ppm and 10 ppm at respectively, at 577 nm (free space wavelength), totalling 80 ppm of coating-induced losses per round trip. The losses are consistent with the known surface roughness. These finesse measurements used an air gap. With ethylene glycol of refrative index 1.44, one would expect finesse less than $1 \times 10^4$, due to enhanced scattering from the rough surface at shorter wavelength, and decreased reflection from the front surface of the dielectric stack. The measured cavity lifetime as discussed in Main Text, Fig. 4 (bottom panel) is just 5 ps. The unaccounted-for losses are most likely due to a combination of chemical impurities in dye and solvent, dye aggregation at high concentrations, and dirt and mechanical damage due to handling mirrors in a standard laboratory rather than a clean room.

The mirrors permit us to reach harmonic trapping frequencies, equivalent to the transverse mode spacing of the optical microcavity, up to $f = 1.7$ THz, or $0.25 \times k_B T$ where $T$ is room temperature, depending on the mirror RoC and cavity length. When $hf$ is of order $k_B T$ finite-size effects become important and the critical photon number for BEC is small. The depth of features is greater than one complete longitudinal free-spectral range. That implies an effective trap depth of greater than $h \times 60$ nm or $10$ $k_B T$, meaning that more than 99.95% of thermal photons are trapped.

It is important to note that the light penetrates the mirrors by an effective number $q_0$ of half-wavelengths. If the true longitudinal mode number (given by the ratio of cavity round trip time to light frequency) is $q$, then the open length of the cavity is $q - q_0$. It is this latter length which enters into calculations for trapping frequency [3]. Also, the effective number density of molecules must be

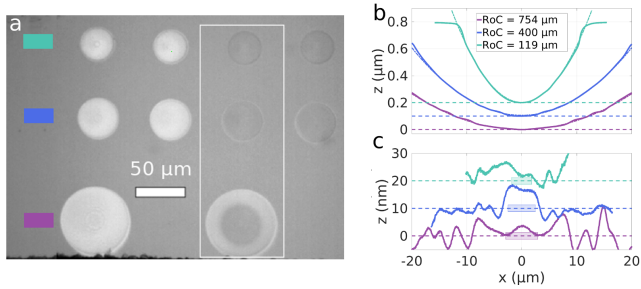

FIG. 1. (a) A white-light microscope image of the mirrors used. (b) Quantitative analysis shows parabolae, and we have used the features with effective radii of curvature 400 $\mu$m. Colours indicate from which row the features are plotted. The white box indicates the column. Features left of the column of (a) are not laser smoothed; figures in the box and to the right have been smoothed. Roughness as shown in (c) is as low as $+0.3$ nm.

re-normalised by a factor $(q - q_0)/q$ (see below).

# A SIMPLIFIED MODEL FOR DYE-MICROCAVITY PHOTONS

We will make use of the simplest version Kirton and Keeling model, which assumes that all dye molecules couple equally to all cavity modes [9, 10], unlike the more recent versions which include spatial [11] and rotational [12] inhomogeneities. The solution of a full quantum master equation can be simplified to a system of rate equations:

$$\frac{\partial n_m}{\partial t} = - \kappa n_m + \\ N_{mol}\left[E(\delta_m)(n_m+1)p_e - A(\delta_m)n_m(1-p_e)\right] \quad (1)$$

$$\frac{\partial p_e}{\partial t} = - \Gamma_\downarrow^{tot}(\{n_m\})p_e + \Gamma_\uparrow^{tot}(\{n_m\})(1-p_e) \quad (2)$$

$$\Gamma_\uparrow^{tot}(\{n_m\}) = \Gamma_\uparrow + \sum_m g_m A(\delta_m)\, n_m \quad (3)$$

$$\Gamma_\downarrow^{tot}(\{n_m\}) = \Gamma_\downarrow + \sum_m g_m E(\delta_m)(n_m+1) \quad (4)$$

where the symbols are: $n_m$, the photon number in the $m$th mode of degeneracy $g_m$; $p_e$ the fraction of excited-state molecules; $N_{mol}$ the total number of molecules; $\Gamma_\uparrow$ the pumping rate; $\Gamma_\downarrow$ the de-excitation rate not including emission into cavity modes (dominated by spontaneous emission into free space). The functions $A(\delta_m)$ and $E(\delta_m)$ are the absorption and emission amplitudes to and from the cavity modes at detuning $\delta_m$ from molecular resonance. $\{n_m\}$ is the set of photon numbers in all modes.

## Thermal-equilibrium limit

The Kennard-Stepanov/McCumber relation imposes that the absorption and emission spectra are related by a principal of detailed balance through rapid vibrational thermalisation of the dye molecules, so that $A(\delta) = E(\delta)\mathrm{e}^{\delta/\mathrm{kT}}$, where $k_B T$ is the thermal energy scale at room temperature. In the limit of low losses $(\kappa, \Gamma_\downarrow \to 0)$, the steady-state solution can be written in the form

$$n_m = \frac{g_m}{\mathrm{e}^{(\delta_m - \mu)/k_B T} - 1}, \quad (5)$$

which is simply the Bose-Einstein distribution. The chemical potential is given by $\exp\left(\mu/k_B T\right) = \Gamma_\uparrow^{tot}/\Gamma_\downarrow^{tot}$.

## Microlaser limit

In a different limit, the model can be compared to a microlaser. First, we simplify the description to just a single cavity mode red-detuned from the molecular resonance, $\delta < 0$, and free space for spontaneous emission. We treat the limit where excitation of the dye molecules by cavity light is negligible, $A \to 0$ while $E$ remains finite. Then equations Eqn. (1) and Eqn. (2) become:

$$\frac{\partial n}{\partial t} = - \kappa n + N_{mol}E(n+1)p_e \quad (6)$$

$$\frac{\partial p_e}{\partial t} = - \left[\Gamma_\downarrow + E(n+1)\right]p_e + \Gamma_\uparrow(1-p_e) \quad (7)$$

These equations are equivalent to standard equations of motion used to describe microlasers [13, 14], which are derived in the approximation that molecular excited-state saturation is negligible, $p_e \ll 1$, an approximation which is typically valid for microcavity experiments not involving polaritons.

These equations are normally written using a parameter $\beta$ which is the fraction of all spontaneous emission which goes into the cavity mode, which is $\beta = E/(E+\Gamma_\downarrow)$ where $E$ is the emission rate evaluated at the cavity detuning. The equations admit analytic, steady-state solution for the population $n$ of the only mode:

$$n = \frac{(\beta\rho - 1) + \sqrt{(1-\beta\rho)^2 + 4\beta^2\rho}}{2\beta} \quad (8)$$

with $\rho = \Gamma_\uparrow N_{mol}/\kappa$ being the normalised pump rate, proportional to pump laser power.

# A COMPLETE MODEL FOR DYE-MICROCAVITY PHOTONS

In this section we describe the full non-equilibrium model which we compare to the data in the Main Text,

and how we ascertain the appropriate microscopic parameters.

The assumption that all photon modes couple equally to all molecules as used in the simplified model of Eqns. (1)–(4) is not valid for many of our experimental parmaeters. Instead, we make use of the updated model of Ref. [11] which takes into account both position-dependent coupling to spatially-varying cavity photon modes and an inhomogeneous excited-state fraction of the dye molecules. Some of us (HJH, RAN and FM) have used precisely this model to describe the phase diagram of multimode photon condensates [15]. We have reached a good quantitative comparison between the experimental data and the model by fixing as many parameters as possible. Here, we justify the values of the remaining parameters.

The first parameter to be set is the cavity loss rate, $\kappa$. From the coherence measurements, in the Main Text Fig. 4 (lower panel), we know that $1/\kappa = 5.2 \pm 0.8$ ps. We fix the value at $1/5$ ps for the remainder of this work unless otherwise stated.

### Light-matter coupling strength

The next parameter to fix is the light-matter coupling strength, $\Gamma(\pm\delta)$. The formalism is stated explicitly in the above-mentioned articles, especially Ref. [15], Eqns. (1)–(3). We note that $\Gamma(\delta)$ is converted to a rate by being multiplied by the molecular density. The correct density $\rho$ to use for our near-planar microcavities is the areal density of molecules, which itself is the bulk molecular concentration multiplied by the length between the mirrors. The molecular concentration is 2.4 mM, giving molecular number density $n_{mol} = 1.4 \times 10^{24} \mathrm{m}^{-3}$. Since the light penetrates the dielectric mirrors, for a cavity of longitudinal mode number $q$, this between-mirror length is about $L_0 = (q - q_0)\lambda^*/2$ where $q_0 \sim 4$ and $\lambda^*$ is the wavelength of light in the medium. In this work we set $q = 8$, and for simplicity we keep $\lambda^* = 400$ nm (equivalent to about 570 nm in free space). Thus $L_0 = 800$ nm, and $\rho = n_{mol}L_0$. The re-absorption rate for a bulk system with in-medium speed of light $c^*$ is $n_{mol}c^*\sigma(\lambda)$ for a measured wavelength-dependent light-scattering cross section $\sigma(\lambda)$. Hence, $n_{mol}\sigma(\lambda)c* = \rho\Gamma(\delta) \Rightarrow \Gamma(\delta) = \sigma(\lambda)c*/L_0$. In the Main Text, the discussion of molecular density uses, for simplicity, the effective concentration of dye molecules taking into account the optical path within the dielectric, written as $\bar{n}_{mol} = n_{mol}(q - q_0)/q$. Finally, conversion between lab-optimised wavelengths $\lambda$ and simulation-optimised detunings $\delta$ is via the definition of a zero-phonon line, where $\Gamma(\delta) = \Gamma(-\delta)$, i.e. the peak-normalised absorption and emission spectra cross, which is at 545 nm (see [16]).

### Molecular decay rate

To match the depth and breadth of the threshold phenomenon in population as a function of pump rate or power, adjusting the fraction of spontaneous molecular de-excitation that goes into creating cavity photons is the most logical step. This fraction is known as the $\beta$ parameter of microlaser physics. Since the intracavity emission is fixed, the only parameter in the model which can set $\beta$ is $\Gamma_\downarrow$, which is expected to be dominated by spontaneous emission into free space. The bulk rate is $\Gamma_\downarrow^{bulk} = 0.25$ ns$^{-1}$. Numerical experiments show $\Gamma_\downarrow = 0.03$ ns$^{-1}$ gives a much more appropriate match to the data. We conclude that spontaneous emission into free space is strongly suppressed by the simple fact that the mirrors subtend a very large solid angle at the centre of the cavity. The mirrors also have a rather broad stop band and a wide angle over which their transmission is very low. Hereafter, and in the Main Text, we fix $\Gamma_\downarrow = 0.03$ nm$^{-1}$.

### Pump-spot size

The only remaining adjustable parameter is the pump spot size. Experimentally, we focus the spot as small as possible, given the numerical aperture available. The lower diffraction limit for our optics is between 1.3 $\mu$m (the diameter of the first zero of the focal Airy disk assuming a uniform beam filling our focussing lens) and 1.7 $\mu$m (based on Gaussian optics for our roughly-Gaussian pump beam). In our simulation, the value is encoded in harmonic oscillator lengths, $l_{ho} = 1.2$ $\mu$m for 1.7 THz harmonic-oscillator mode spacing. In the simulation, we assume an elliptical spot, whose major axis is $\sqrt{2}$ times larger than its minor axis, since the beam is incident on the cavity at approximately $45°$ to pass through the dielectric mirror efficiently. From one experimental run to the next the pump spot may be re-aligned and re-focused to the smallest achievable spot size, with precision roughly equivalent to the depth of focus, i.e. we reliably achieve spot sizes of $1$–$2l_{ho}$. We focus as small as possible to minimise pumping highly-excited transverse modes of an secondary longitudinal mode of the cavity. We detect those modes in experimental spectra, and can pass lasing threshold if we do not carefully align and focus our pump beam.

In Fig. 2 we see results of a simulation of thermalisation of light (far below threshold pump rate). The parameters are as described above, but we vary both cavity cutoff wavelength and pump spot size, for three possible cavity lifetimes. The Boltzmann distribution is then fitted to the mode populations and an effective temperature extracted. For the longest cavity lifetimes, there is a substantial region (large pump spot, 560–580 nm cutoff)

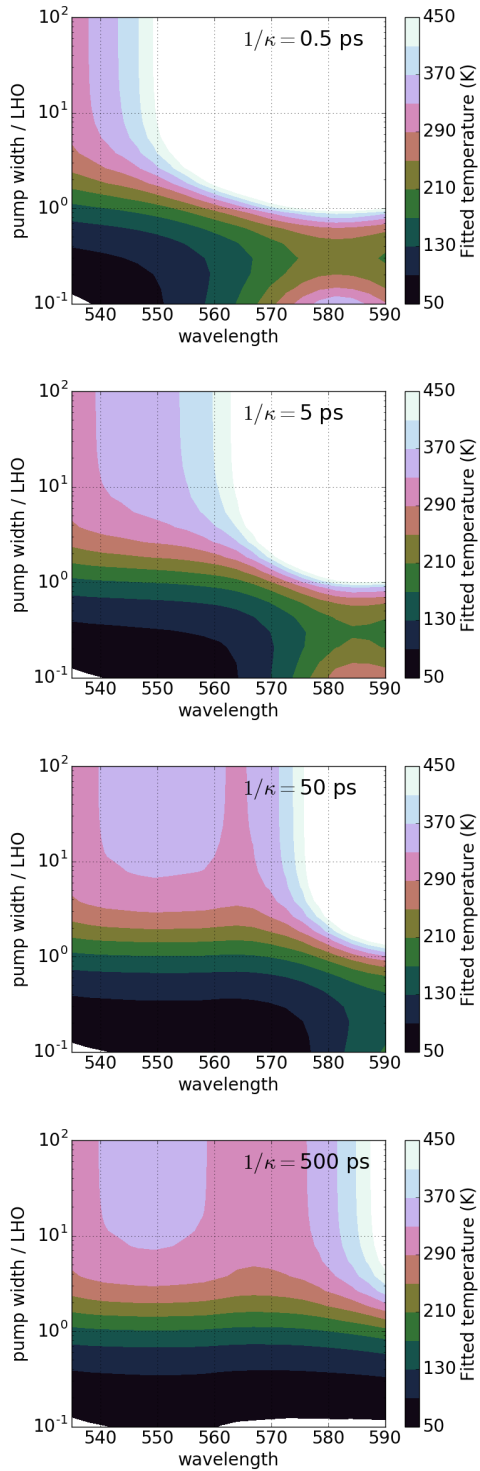

FIG. 2. Effective temperature (see side bar for scale) derived by fitting simulated mode populations far below threshold pump rate, as a function of pump spot size and cavity cutoff wavelength. The inverse of the spontaneous emission lifetime is set to $1/\Gamma_\downarrow = 32$ ns as determined from the width of the threshold feature. Top to bottom, the cavity lifetime is increased.

over which the temperature matches room temperature, independent of the precise parameter values. This represents true thermal equilibrium, and requires at least 20 re-absorption events per cavity loss. As the cavity lifetime is reduced, the region covers a smaller range of cutoff wavelengths, and even disappears for the lifetimes we believe describe our experiments, 5 ps. For that lowest lifetime, there is no true thermal equilibrium but that re-absorption and emission events have significantly redistributed the cavity light from the emission spectrum of the dye towards the thermal equilibrium distribution. In the Main Text, Fig. 1(c) we find effective temperatures of 170±20 K for 560 nm cutoff wavelength, suggesting pump spot very close to the diffraction limit, around 1 $l_{ho}$. Our experiment therefore is in a non-equilibrium regime (in contrast to say Ref. [17]), but nonetheless a near-equilibrium description with an adjustable effective temperature is appropriate.

The simulations here and in the Main Text are run using a total of 15 modes across 5 non-degenerate energy levels. We have run example simulations using 28 modes over 7 levels, and found no significant differences in the results.

## NON-EQUILIBRIUM DEFINITIONS OF CONDENSATION

There is no unique recipe for determining the threshold for condensation, especially in finite-sized systems and non-equilibrium. A common starting point is the Penrose-Onsager criterion [18] which can be interpreted as stating that condensation occurs when the occupation of a mode of the multi-particle system has an occupancy which grows linearly with the total number of particles in the system. They call this specifically *Bose-Einstein* condensation. Their definition does not strictly distinguish what we are labelling as single-mode Bose-Einstein condensation from multimode condensation. It also cannot be trivially mapped onto finite-sized systems as the notion of extensivity is only valid in the thermodynamic limit.

Let us now review several alternative operational definitions of threshold from a variety of physical implementations (laser physics, atomic gases, general statistical physics) with the aim of applying them to our finite-size, non-equilibrium system with multiple modes and variable total particle number, which capable of showing multimode condensation.

A. Generalised condensation in a specific mode labelled $i$ occurs when the number of particles in that mode $n_i$ is a finite and constant fraction of the total particle number $n_{tot}$, but the populations in non-condensed (or depleted) modes are not proportional, as the total particle number goes to infinity: $\lim_{n_{tot} \to \infty} n_i/n_{tot} =$

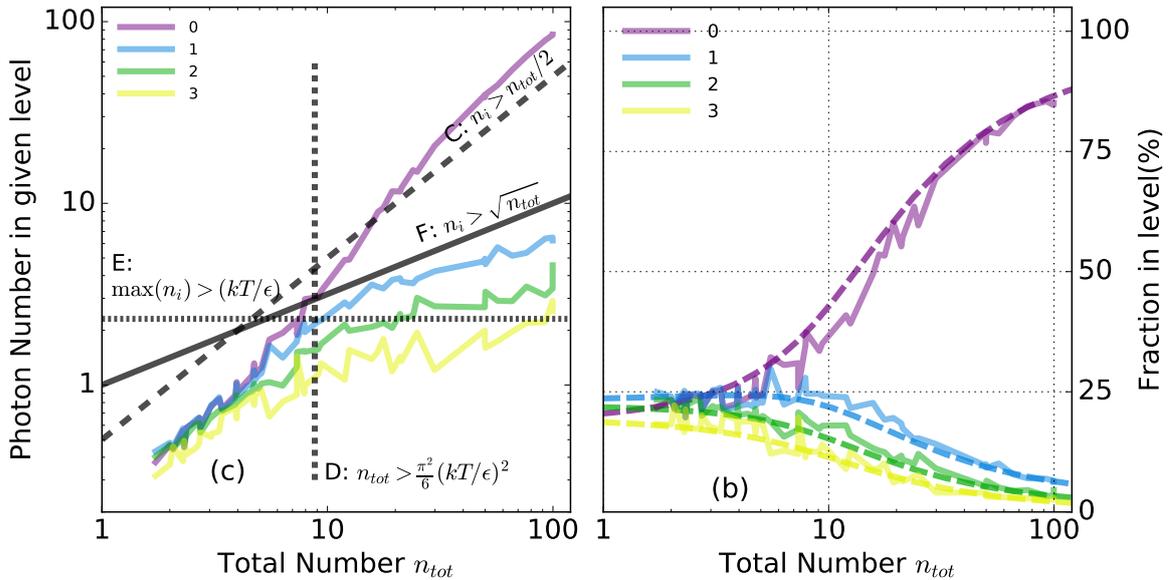

FIG. 3. Criteria for determining the condensation threshold: see text for descriptions. Criteria A. and B. dot not allow us to determine threshold, and G. is not relevant to our multi-mode system. Left panel: number in each mode as a function of total photon number, together with labelled criteria. A temperature fit of 170 K was used, with energy level spacing $h \times 1.7$ THz. Right panel: the same data, as a fraction, showing clearly that an increasing fraction of photons go into the ground mode for increasing photon number. The non-equilibrium model results are shown as dashed lines.

$\mathcal{O}(1)$. This criterion is discussed in Ref. [19] in a very general context and holds for both Bose-Einstein and multimode condensation, regardless of notions of thermal equilibrium. While it makes clear whether condensation happens or not, because it is only applicable in the asymptotic limit of large particle number, a well-defined threshold criterion cannot be derived from it.

B. A linear-growth condition for BEC is when criterion A. holds for one and only one mode $i = 0$, that being the lowest energy mode. This holds for non-equilibrium systems but only in the asymptotic limit of large $n_{tot}$ and so cannot be used to determine threshold. This is a specialised version of and is used in, for example, Ref. [20].

C. If more than half of all particles are in one state, then that must be the only condensed state. If that mode is the lowest energy mode, one can then say that the mode is Bose-Einstein condensed. Since it holds in a finite system it can be used to define threshold: $n_i/n_{tot} > 1/2$. Criteria A. and B. are necessary but not sufficient conditions for this criterion; this is a very strict criterion, used by Ref. [21].

D. In trapped, conservative, thermal-equilibrium Bose gases (e.g. cold atomic vapours) BEC is considered to occur when $n_{tot} > \lim_{n_{tot} \to \infty} n_{tot} - n_0$, where the total population exceeds the saturated excited-mode population in the infinite particle-number limit [22].

For simple trapping potentials this limit can be calculated analytically. For example, in a 2D trap of angular frequency $\Omega$ and energy $\hbar\Omega = \epsilon$, relevant to our experiments, this criterion becomes $n_{tot} > \left(\pi^2/6\right)\left(k_B T/\epsilon\right)^2$, for one polarisation state. This criterion is valid only for thermal equilibrium in a well-known potential energy landscape..

E. In criterion D., at threshold the ground state population is well defined. To leading order one finds $n_0 > k_B T/\epsilon$ [10] as an alternative criterion suitable for finite-sized systems. Despite being derived from equilibrium considerations, Kirton and Keeling [10] use it even away from equilibrium, with generalised condensation occuring when at least one mode (not necessarily the ground mode) has this population.

F. A consequence of the same analysis is $n_0 = n_{tot}^{1/\alpha}$ at the threshold, with $\alpha > 1$ being the dimensionality. For our system, $\alpha = 2$ so we find a criterion of $n_0 > \sqrt{n_{tot}}$ [10, 22]. While this is strictly an equilibrium criterion, it still makes sense out of equilibrium, generalised to $n_i > \sqrt{n_{tot}}$. When there are few particles, few modes may condense. With many particles, many, even all, modes may condense. This $\mathcal{O}(n^{1/2})$ criterion neatly separates saturated or depleted states with $n_i = \mathcal{O}(n^0)$ from condensed modes where $n_i = \mathcal{O}(n^1)$.

G. Mode occupancy of one, $n_i > 1$, is considered a threshold for microlasers by some authors (e.g.

Ref. [13]). This is a non-equilibrium condition but based on a single mode. We consider condensation (as opposed to lasing) to be a phenomenon which occurs in multi-mode systems, this criterion is not useful to us here, except in comparison to the micro-laser model of Eqn. (8).

It is possible to invent other *ad hoc* criteria for threshold, but robust criteria which can be applied in the presence of experimental noise are not obvious.

In Fig. 3 we see the same data as in the Main Text Fig. 1(c). We determine threshold by comparison with the criteria listed here. Only criteria D. and E. make use of the temperature extracted from the fits to the Bose-Einstein distribution. If the criteria all represented the same physical interpretation of threshold, we would expect them to pass through one unique point. They do not, implying that the criteria are logically distinct. Despite the system being out of equilibrium, the two equilibrium criteria D. and E. (applied with $T = 170\,K$ as fitted) coincide closely with the data to within experimental noise. Criterion F. almost coincides. The coincidence of all three criteria bracket the data, at a total photon number of $8 \pm 1$ with $\sqrt{8}$ photons in the ground state. This is the figure we state for threshold in the Main Text. Only C. gives a significantly different value of threshold, at around 16 photons. We consider this latter criterion overly strict for establishing BEC unless there is some ambiguity which the other criteria cannot be used to resolve.

Despite the imperfect saturation of excited states due to the deviation from thermal equilibrium, according to criterion F., only the ground state goes above threshold within the range of data taken here. All other modes remain below threshold or depleted. The most common operational concept of BEC as large-scale occupation of the ground state alone is then clearly applicable to our system, despite is finite size and imperfect equilibration.

## FITTING THE BOSE-EINSTEIN AND MICROLASER DISTRIBUTIONS AND CALIBRATING PHOTON NUMBER

To extract a temperature we fit the Bose-Einstein distribution to the data, as shown in the Main Text, Fig. 1(c). The procedure is a hierarchical least-squares fit using log-spaced residuals to account for the large dynamical range of the data. The function fitted is the Bose-Einstein distribution, summed over the first five lowest energy levels (15 modes). In an outer optimisation covering the complete data set to be fitted, the temperature $T$ and a number calibration scale $N$ are selected. In an inner loop, for each total particle number, the chemical potential is varied to minimise the residuals between the data and the distribution across the 5 modes,

using the specified value of $T$. The temperature $T$ and calibration $N$ are varied to globally minimise the total residuals. Not all data are used for the fit. Far above threshold, the excited state populations do not saturate which introduces a bias to the fitted temperature. We truncate the data at around $n_{tot} = 60$ in this case, as represented by the termination of the solid lines in the Main Text, Fig. 1(c). Varying the truncation point varies the temperature by approximately 20 K, hence we quote $170 \pm 20$ K for the fitted temperature.

To fit the microlaser distribution, Eqn. (8) we fit to the ground mode population alone, as a function of pump power. There are three parameters: the spontaneous emission fraction $\beta$, the number calibration $N$ and a scaling $G$ between nominal pump laser power and the pump rate of the molecules in units of the cavity lifetime. This latter parameter is of no fundamental importance.

We calibrate the spectrometer with a known pump power laser beam, but the coupling from intra-cavity photons to the spectrometer detection is uncertain, hence the need for the number calibration parameter $N$ in the fits. The fits to the two different distribution yield similar calibration values, of approximately 0.3, implying that there we have unaccounted-for but not-unreasonable losses of about 70% of our light between cavity and detector. A large part of the difference may come from the mirror transmission, for which we use the manufacturer's calculated specification interpolated to our required wavelengths rather than a direct measurement. In practice, having used fitting to check the value of $N$, we fix one value and then re-fit both models.

## POLARISATION STATES

In the thermal equilibrium model we take the degeneracy of the $i$th excited level to be $i+1$ as appropriate for a symmetric 2D harmonic oscillator potential. In doing so, we assume that there is only one polarisation mode for the light. In reality there are two. These two polarisations introduce an ambiguity to the experimental observation of threshold: if one mode condenses but not the other, how could we tell without polarisation-sensitive detection?

The polarisation of photon BECs has only very recently been studied theoretically [12] and experimentally [23]. The main conclusion of the theoretical modelling is that the polarisation degree of freedom thermalises through rotational diffusion of the dye molecules, which is a much slower process than the energetic or spatial thermalisation. The experiments show that while the BEC (the ground state when condensed) is mostly linearly polarised along the pump axis, the other modes remain unpolarised. This evidence suggests that the two polarisation degrees of freedom interact only weakly with each other, especially in our system with a rather short

cavity lifetime (5 ps) compared to the rotation diffusion time constant (about 2 ns).

In our experiments we place a polarising filter in front of the spectrometer used for detection, aligned to maximise transmission above threshold photon number. Since the unobserved, orthogonal polarisation component does not interact with the observed component, we simply make the assumption that the polarisation degeneracy is unity for our calculations. This is standard practice in photon BEC for measuring, for example, photon-photon correlations [24] or coherence [2, 17].

## DRIVEN-DISSIPATIVE THEORY OF MICROLASER COHERENCE

The well-established master equation for the molecules and photons among the various cavity modes is numerically and analytically difficult to solve, so we will make a single-mode approximation, as done in Ref. [10]. The density operator can be solved for in a basis using photon numbers and excited molecule numbers. From the number–off-diagonal elements, the first-order correlation function $g^{(1)}(\tau)$ as a function of relative time can be calculated. A time-evolution equation for those off-diagonal elements can be extracted from the full master equation, and solved through recursion of higher and lower photon numbers, assuming that the molecular state is independent of the photon state. Such an approximation is likely to be valid except in the limit of very large photon numbers, and also in the multimode condensate regimes.

In the steady state, the first-order correlation function $g^{(1)}(\tau)$ is approximately exponential as a function of the absolute value of $\tau$, but not necessarily exactly (so the spectrum may deviate from Lorentzian). Nonetheless, one can define a coherence time from the decay of $g^{(1)}$. Numerical solution of the equations is intractable for experimentally appropriate molecule numbers (around $10^6$), but assuming molecule and photon states are separable, mean-field results can be obtained.

The model is similar to Eqn. (6) and Eqn. (7), but retaining the absorption coefficient $A$. One solves for the average excited-state fraction of molecules $p_e$ and obtains:

$$N_{mol}p_e = \frac{\Gamma_\uparrow^{tot} N_{mol}}{\Gamma_\downarrow^{tot} + \Gamma_\uparrow^{tot}} = \frac{[\kappa - A(\delta)N_{mol}]\langle n \rangle}{E(\delta)[\langle n \rangle + 1] + A(\delta)\langle n \rangle} \quad (9)$$

where $\Gamma_\uparrow^{tot} = \Gamma_\uparrow + A(\delta)\langle n \rangle$ and $\Gamma_\downarrow^{tot} = \Gamma_\downarrow + E(\delta)(\langle n \rangle + 1)$, and $\langle n \rangle$ is the mean number of photons in the mode of detuning $\delta$ relative to the zero-phonon line.

Then the Lorentzian linewidth $\Gamma^T$, equivalent to the inverse coherence time $1/\tau_c$ is given by:

$$\Gamma^T = \frac{1}{2}\left[\kappa + A(\delta)N_{mol}(1 - p_e) - E(\delta)N_{mol}\,p_e\right]. \quad (10)$$

The photon absorption rate $A(\delta)$ in the limit of weak molecular excitation $p_e \ll 1$ is related directly to the measured absorption cross-section $\sigma(\lambda)$ by $N_{mol}A = n_{mol}\sigma(\lambda)c^*$. Here $c^*$ is the speed of light in the medium, $\delta = 2\pi c\left(\frac{1}{\lambda_{ZPL}} - \frac{1}{\lambda}\right)$ is the detuning from the molecular zero-phonon line and $n_{mol}$ is the effective number density of molecules (given by the true molecular density times the fraction of the light which is in the open part of the microcavity not within the mirrors, $(q - q_0)/q$). Thus in the limit of weak pumping and few photons we find a simple expression for the decoherence rate (inverse coherence time) which contains experimentally measurable quantities:

$$\frac{1}{\tau_c} = \frac{1}{2}\left[\kappa + n_{mol}\sigma(\lambda)c^*\right]. \quad (11)$$

Decoherence is simply caused by photon loss, either by emission from the cavity or by re-absorption, which is the mechanism by which photon thermalisation occurs. In the limit of large photon numbers, it can be shown that the coherence time is proportional to photon number [10].

The theory matches qualitatively well to the experimental data for small and moderate photon numbers, with the only adjustable parameter being $\kappa$. Photon number calibration uses the same method and value as the Main Text Fig. 2. We assume here that the dye concentration is 2.4 mM, which is slightly higher than the expected value but within uncertainty, which is perhaps a factor 2 [1]. The dye is changed between data sets and concentration is not always constant which means that the the data in the two panels of Main Text Fig. 4 show small but statistically significant differences.

---

* Correspondence to r.nyman@imperial.ac.uk




\end{document}